\documentclass[%
 letterpaper,twocolumn,english,
 superscriptaddress,
showpacs,preprintnumbers,
 amsmath,amssymb,
 aps,
 prc,
floatfix,
]{revtex4-1}

\usepackage[usenames]{color}
\usepackage{amsmath}
\usepackage{amssymb}
\usepackage{longtable}  
\usepackage{graphicx}
\usepackage[dvipdfm,ps2pdf,unicode=true,bookmarks=false,breaklinks,pdfborder={0
    0 0},backref=false,colorlinks=true]{hyperref}  
\hypersetup{pdftitle={Double Spin Asymmetries of Inclusive Hadron
    Electroproductions from a Transversely Polarized
    He-3 Target},pdfauthor={Y.X. Zhao , et. al. (The Jefferson Lab Hall
    A Collaboration)},pdfsubject={v1: nucl-ex,
    hep-ex},pdfkeywords={Transverse DSA Inclusive hadron SIDIS He-3
    $A_\text{LT}$ TMD},linkcolor=DarkBlueCite, citecolor=DarkBlueCite,
  urlcolor=DarkBlueCite}  
\usepackage[hyphenbreaks]{breakurl}

\makeatletter

\usepackage{dcolumn}
\usepackage{bm}

\definecolor{DarkBlueCite}{rgb}{0.1,0.0,0.5}

\makeatother

\begin{document}



\author{Y.X.~Zhao}\email[Corresponding author: ]{yxzhao@jlab.org}
\affiliation{University of Science and Technology of China, Hefei 230026, People's Republic of China}
\author{K.~Allada}
\affiliation{Massachusetts Institute of Technology, Cambridge, MA 02139}
\affiliation{Thomas Jefferson National Accelerator Facility, Newport
  News, VA 23606}  
\author{K.~Aniol}
\affiliation{California State University, Los Angeles, Los Angeles, CA 90032}
\author{J.R.M.~Annand}
\affiliation{University of Glasgow, Glasgow G12 8QQ, Scotland, United Kingdom}
\author{T.~Averett}
\affiliation{College of William and Mary, Williamsburg, VA 23187}
\author{F.~Benmokhtar}
\affiliation{Carnegie Mellon University, Pittsburgh, PA 15213}
\author{W.~Bertozzi}
\affiliation{Massachusetts Institute of Technology, Cambridge, MA 02139}
\author{P.C.~Bradshaw}
\affiliation{College of William and Mary, Williamsburg, VA 23187}
\author{P.~Bosted}
\affiliation{Thomas Jefferson National Accelerator Facility, Newport
  News, VA 23606}
\author{A.~Camsonne}
\affiliation{Thomas Jefferson National Accelerator Facility, Newport News, VA 23606}
% ...
\collaboration{The Jefferson Lab Hall A Collaboration}
\noaffiliation

\title{Double Spin Asymmetries of Inclusive Hadron 
  Electroproductions from a Transversely Polarized $^3\rm{He}$ Target} 


\begin{abstract}
We report the measurement of beam-target double-spin 
asymmetries ($A_\text{LT}$) in the inclusive production of identified 
hadrons, $\vec{e}~$+$~^3\text{He}^{\uparrow}\rightarrow h+X$, 
using a longitudinally polarized 5.9 GeV electron beam 
and a transversely polarized $^3\rm{He}$ target. 
Hadrons ($\pi^{\pm}$, $K^{\pm}$ and proton) were detected at 
16$^{\circ}$ with an average momentum
$<$$P_h$$>$=2.35 GeV/c and a transverse momentum ($p_{T}$) coverage from 0.60 to 0.68 GeV/c. 
Asymmetries from the $^3\text{He}$ target 
were observed to be non-zero for $\pi^{\pm}$ production when the 
target was polarized transversely in the horizontal
plane. The $\pi^{+}$ and $\pi^{-}$ asymmetries have opposite signs, analogous
to the behavior of $A_\text{LT}$ in semi-inclusive deep-inelastic scattering.

\end{abstract}


\maketitle


\section{Introduction}

Understanding the spin structure of the nucleon remains an important goal of research 
in modern hadronic physics. Beam-target double-spin asymmetries (DSA) have
been used as a powerful tool in polarized lepton-nucleon deep-inelastic scattering (DIS)
experiments to extract polarized parton distributions and quark-gluon correlations \cite{JPreview}. 
Earlier efforts have been focused mainly on the longitudinal
spin structure $g_{1}$. Recently, with transversely polarized nucleons,
DSAs were used to investigate the $g_{2}$ structure functions, which involve 
twist-3 effects. More recently, a measurement of DSA with a transversely polarized nucleon
($A_\text{LT}$) in a semi-inclusive deep-inelastic scattering (SIDIS) experiment has provided
access to the transverse-momentum-dependent parton distribution functions $g_{1T}(x,k^2_t)$, which
are related to quark spin-orbit correlations \cite{Huang2012}. In this paper, a measurement of $A_\text{LT}$ in
a less explored reaction, $\vec{e}~$+$~\text{N}^{\uparrow}\rightarrow h+X$, in which a single hadron is 
detected in the final state, is presented.

The mechanism of inclusive hadron photoproduction was studied in \cite{PhysRevD.58.054007, PhysRevD.61.034014}. 
The production of hadrons arises mainly from four types of processes: fragmentation processes,
direct processes, resolved photon processes and soft contributions. 
Fragmentation processes have quarks and gluons produced in short-range reactions followed
by fragmentation at long distances of either a quark or a gluon to produce the observed hadron.
Direct processes occur when the hadron
is produced in a short-range reaction via a radiated gluon giving a quark-antiquark
pair, one of which joins the initial quark to produce the hadron. 
Resolved processes are contributions
in which photons fluctuate into a quark-antiquark pair, which then interact with the partons of the target.
Soft contributions are described by the vector meson dominance (VMD) approximation, which is a way to represent
the hadronic components of the photon as they enter into soft processes.

In the collinear factorization framework, $A_\text{LT}$ in inclusive hadron production 
is an observable associated with twist-3 effects. It can have twist-3 contributions from 
both the parton distributions inside the polarized nucleon and the parton fragmention 
into final state hadrons. By measuring $A_\text{LT}$, one has the opportunity to investigate 
the ``worm-gear"-type function $\widetilde{g}(x)$ \cite{Alt_paper,PhysRevD.81.054008} as well 
as the role of quark-gluon-quark correlations in the nucleon and twist-3 effects in the 
fragmentating hadron. The $\widetilde{g}(x)$ is defined as an integration \cite{Alt_paper} 
over $k^{2}_t$ of $g_{1T}(x,k^2_t)$, which can be accessed by $A_\text{LT}$ measurements in a SIDIS process \cite{Huang2012}.
Furthermore, it has been proposed that $\widetilde{g}(x)$ and quark-gluon-quark 
correlations are responsible for DSAs of inclusive jet (or hadron) production in 
polarized nucleon-nucleon reactions and lepton-nucleon 
reactions in \cite{PhysRevD.84.034046, PhysRevD.86.114020}.

In this paper, we report a measurement of beam-target double-spin asymmetries in inclusive 
charged-hadron production using a longitudinally polarized electron beam scattered from a transversely 
polarized $^3\rm{He}$ target. The measured asymmetry is defined as
\begin{equation}
A_{\text{LT}} = \frac{1}{|P_BP_{target}|}\frac{d\sigma^{\uparrow \rightarrow}-d\sigma^{\downarrow \rightarrow}}{d\sigma^{\uparrow \rightarrow}+d\sigma^{\downarrow \rightarrow}},
\label{eqn_DSA_definition}
\end{equation}
where $d\sigma^{\uparrow (\downarrow) \rightarrow}$ is the differential
cross-section for beam helicity + (-) in a certain target spin direction.
$P_B$ is the beam polarization and $P_{target}$ is the target polarization.
Figure 1 shows the kinematical configuration in the laboratory coordinate
system of the measurement. $\phi_{s}$ is the azimuthal angle between the target spin
direction $\vec{S}$ and the ``hadron plane" which is formed by the incoming
electron and the outgoing hadron. The spin-dependent part of the cross-section
is proportional to the term $\lambda_e \vec{S}\cdot\vec{p}_{T}$ ($p_{T}=\sqrt{p_x^2+p_y^2}$, 
the transverse momentum of the outgoing hadron), which gives rise to a cos($\phi_s$)
modulation in the definition of the asymmetry \cite{Alt_paper}.
In order to form the parity-even structure by using the spin of the nucleon and the 
momentum of outgoing hadron, cos($\phi_s$) is the only modulation considered in
the current theoretical framework \cite{Alt_paper}.
Hence, the asymmetry can be written as
\begin{equation}
A_{\text{LT}} = A^{\cos(\phi_{s})}_{\text{LT}}\cos(\phi_{s}).
\label{eqn_DSA_cos_definition}
\end{equation}
The produced hadrons were detected in a high-resolution spectrometer
(HRS) \cite{Alcorn2004} at a central angle of 16$^{\circ}$ on the beam left side with a central
momentum of 2.35 GeV/c, a momentum acceptance of $\pm$4.5\% and solid angle acceptance of 6 msr.
The data were collected using a singles
trigger during the E06-010 experiment \cite{Qian2011,Huang2012,PhysRevC.89.042201,PhysRevC.90.055201} 
in Hall A at Jefferson Lab.

\begin{figure}[ht]
\includegraphics[width=70mm]{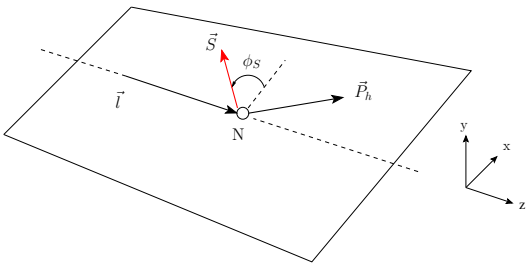}
\caption{(Color online) Kinematical configuration in the laboratory
coordinate system for the $\vec{e}N^{\uparrow}\rightarrow hX$
process. $\vec{l}$ ($\vec{P}_h$) represents the momentum direction of the incident
electron (produced hadron), and $\vec{S}$ is the spin vector of the nucleon.
During the experiment, the target spin was oriented 
in $\phi_{s}=0^{\circ}(+x), 90^{\circ}(+y), 180^{\circ}(-x), 270^{\circ}(-y)$ directions.}
\label{fig_1}
\end{figure}

\section{Experiment}

A polarized 5.9 GeV electron beam with an average current of 12 $\mu$A was
provided by the CEBAF accelerator during the experiment. Polarized electrons
were excited from a strained superlattice GaAs photocathode by a circularly polarized
laser \cite{PhysRevSTAB.10.023501} at the injector. The average beam polarization was 
(76.8 $\pm$ 3.5)$\%$, which was measured periodically by M{\o}ller polarimeter \cite{Alcorn2004}. 
The beam helicity was reversed at 30 Hz by flipping the laser polarization. During
the E06-010 experiment, the sequence for beam helicity states followed a quartet
structure, + - - + or - + + -, randomly to reduce the systematic bias between 
the two helicity states. Due to a beam-charge feedback system \cite{Androić201159}, 
the beam-charge asymmetry between the two helicity states was kept at less 
than 150 ppm per 20 minutes and less than 10 ppm for the entire experiment \cite{Huang2012}.

The ground state of the $^3\rm{He}$ nuclear wavefunction is dominated by the S-state, in
which the proton spins cancel each other and the nuclear spin is carried
by the neutron \cite{Bissey2001}. About 10 atm of $^3\rm{He}$ gas was filled in a 40 cm-long cylindrical
aluminiosilicate glass cell and $^3\rm{He}$ nuclei were polarized by spin-exchange optical
pumping of a Rb-K mixture \cite{Babcock2005,E06010_target}. Three pairs of Helmholtz coils were used in the experiment
to orient the magnetic holding field transversely or vertically with respect to the electron beam. 
For each orientation, the spin direction of $^3\rm{He}$ nuclei was flipped 
every 20 minutes through adiabatic fast passage. Nuclear magnetic resonance measurements, 
calibrated by the electron paramagnetic resonance method \cite{EPR}, were performed to monitor the target polarization
while the target spin direction was flipped. An average in-beam target polarization of
(55.4 $\pm$ 2.8)$\%$ was achieved during the experiment.

The HRS detector package was configured for hadron detection. The trigger was formed by 
the coincidence signal between two scintillator planes which were about 2 meters apart. 
Four detectors were used for particle identification: 
1) a threshold CO$_2$ gas Cerenkov detector for electron identification,
2) a threshold aerogel Cerenkov detector for pion identification,
3) a ring imaging Cerenkov (RICH) detector for $\pi^{\pm}$, $K^{\pm}$, and proton identification \cite{PhysRevC.89.042201, youcai_thesis},
4) two layers of lead-glass calorimeter for electron-hadron separation.
Contaminations were well controlled and studied carefully in \cite{PhysRevC.89.042201}.

\section{Data Analysis}

For each target spin direction, the selected data samples were separated into two groups by beam helicity states.
These two groups were treated as a ``local pair". The final beam-target double-spin asymmetry $A_{\text{LT}}$ 
was extracted by summing over all ``local pair" measurements.

A small amount of N$_2$ gas, present in the target cell to reduce depolarization \cite{Alcorn2004}, diluted the measured
$^3\rm{He}$ asymmetry and was corrected by the nitrogen dilution factor defined as
\begin{equation}
f_{\text{N}_{2}}=\frac{\rho_{\text{N}_{2}}\sigma_{\text{N}_{2}}}{\rho_{^{3}\rm{He}}\sigma_{^{3}\text{He}}+\rho_{\text{N}_{2}}\sigma_{\text{N}_{2}}},
\label{dilequation}
\end{equation}
where $\rho$ is the density of the gas in the production target cell
and $\sigma$ is the unpolarized inclusive hadron (pion, kaon and proton) 
production cross section. The ratio of
unpolarized cross sections $\sigma_{\text{N}_{2}}$/$\sigma_{^3{\rm{He}}}$
was measured in dedicated runs on targets filled with
known amounts of unpolarized $\text{N}_{2}$ or $^3{\rm{He}}$ gas.
The $f_{\text{N}_{2}}$ in this experiment was determined to be less than 10$\%$.

The overall systematic uncertainty in the experiment was small due to frequent 
target-spin and beam-helicity flips. The false asymmetry due to luminosity fluctuations
was less than 0.07\% and was confirmed by measuring the beam-target double-spin asymmetry in
the inclusive (e,e') DIS reaction with the target polarized in the $\pm{y}$ direction, 
in which the asymmetry vanishes due to parity and time-reversal symmetry. Systematic 
uncertainties due to contaminations were estimated to 
be less than 0.02\% for pion, kaon and proton measurements. 
In addition, there was an overall 5\% systematic 
uncertainty, relative to the asymmetries, from both beam and target polarizations.
For the kaon and proton measurements, as described in \cite{PhysRevC.89.042201},
there were two additional sources of systematic uncertainties associated with the RICH detector:
1) the value of the cut on the number of hits in the RICH detector;
2) detector inefficiencies.
The first contribution was determined to be 
$<$15\% for $K^{\pm}$, and $<$3\% for protons, relative to the statistical 
uncertainties. The second contribution was determined to 
be $<$7\%, $<$3\%, and $<$1\%, relative to the statistical
uncertainties, for $K^{+}$, $K^{-}$ and protons, respectively. 

\begin{figure}
\begin{center}
\includegraphics[width=85mm]{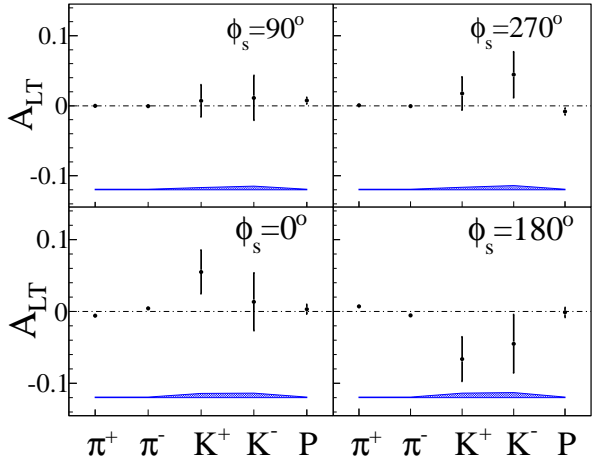}
\end{center}
\caption{(Color online) Beam-target double-spin asymmetries $A_\text{LT}$ for 
$\pi^{\pm}$, $K^{\pm}$ and proton production from $^3\rm{He}$ for different $\phi_{s}$.
}
\label{he3_result_plot}
\end{figure}

\begin{figure}
\begin{center}
\includegraphics[width=85mm]{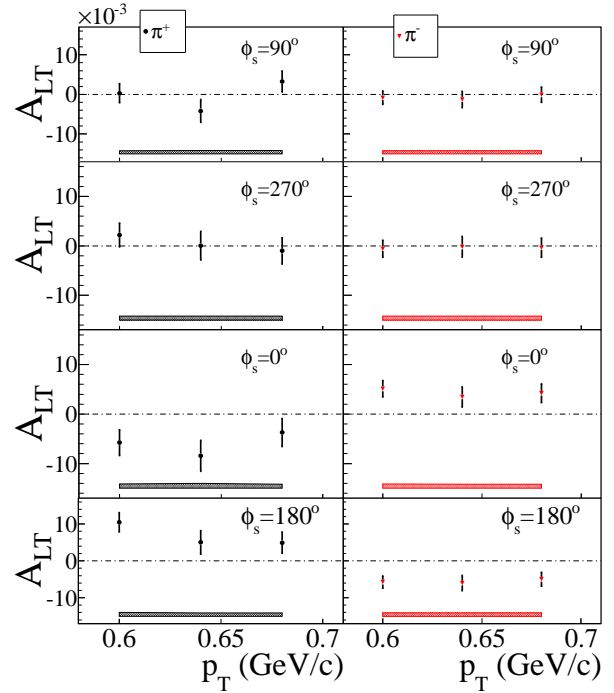}
\end{center}
\caption{(Color online) Beam-target double-spin asymmetries $A_\text{LT}$ for $\pi^{\pm}$ production 
from $^3\text{He}$ as a
function of $p_{T}$ for different $\phi_{s}$. The left column is for the $\pi^{+}$ data,
the right column is for the $\pi^{-}$ data.}
\label{pion_3_bin_plot}
\end{figure}

\section{Results}

\begin{figure}
\begin{center}
\includegraphics[width=85mm]{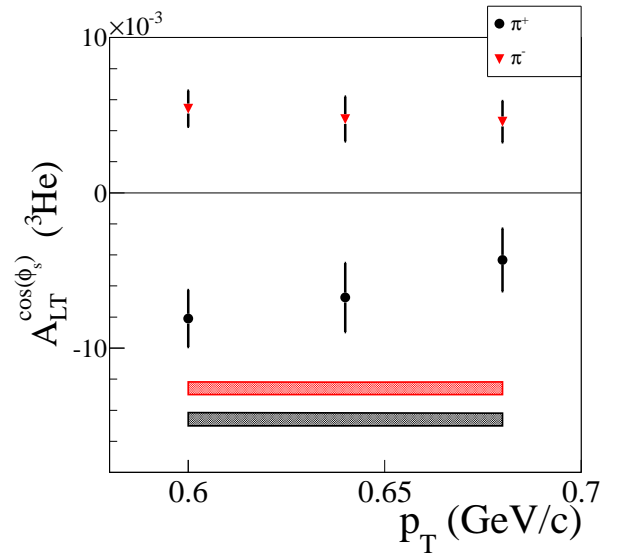}
\end{center}
\caption{(Color online) Beam-target double-spin asymmetries $A_\text{LT}^{\cos(\phi_{s})}$ 
for $\pi^{\pm}$ production from $^3\rm{He}$ as
a function of $p_{T}$.
The red (top) band is the systematic uncertainty band
for $\pi^{-}$, and the black (bottom) band is the systematic uncertainty band for $\pi^{+}$.
}
\label{he3_pion_result_plot}
\end{figure}

The final $A_{\text{LT}}$ results from $^3\rm{He}$ are shown for different hadron
species in Figure \ref{he3_result_plot} and tabulated in Table \ref{he3_result_all_species_table}.
The error bars represent the statistical uncertainties. Experimental systematic uncertainties, combined
in quadrature from different sources, are shown as a band. For $\phi_{s}=90^{\circ}$ and $270^{\circ}$, the asymmetries from 
pions and kaons are consistent with zero within the experimental uncertainties ($\sim 1 \times 10^{-3}$ level for the pion measurement). 
For $\phi_{s}=0^{\circ}$ and $180^{\circ}$, the sign of the asymmetry is flipped when the target spin direction is reversed.
Pion data were also analyzed in three $p_{T}$ bins. The results are shown in Figure \ref{pion_3_bin_plot}. 
The asymmetries for $\phi_{s}=0^{\circ}$ and $\phi_{s}=180^{\circ}$ were
combined together to obtain $A_\text{LT}^{\cos(\phi_{s})}$. The combination was weighted by the statistical uncertainties
of the asymmetries. The final $p_{T}$-dependent $A_\text{LT}^{\cos(\phi_{s})}$ asymmetries for $\pi^{\pm}$ production from $^3\rm{He}$
are shown in Figure \ref{he3_pion_result_plot} and tabulated in Table \ref{he3_pion_result_table}.

\begin{table}
\begin{tabular}{|c|c|c|}\hline
$<$$p_T$$>$       &    $\pi^{+}$       &     $\pi^{-}$                   \\
(GeV/c)           &   ($A_\text{LT}^{\cos(\phi_{s})}\pm$ Stat.$\pm$ Sys.) & ($A_\text{LT}^{\cos(\phi_{s})}\pm$ Stat.$\pm$ Sys.)         \\ \hline
0.60              & -0.0081 $\pm$ 0.0018  $\pm$ 0.0009     &  0.0054 $\pm$0.0012   $\pm$0.0008     \\
0.64              & -0.0067 $\pm$ 0.0022  $\pm$ 0.0008     &  0.0048 $\pm$0.0014   $\pm$0.0008        \\
0.68              & -0.0043 $\pm$ 0.0020  $\pm$ 0.0008     & 0.0046  $\pm$0.0013   $\pm$0.0008    \\   \hline
\end{tabular}
\caption{Tabulated results of $p_{T}$ dependent $A_\text{LT}^{\cos(\phi_{s})}$ for $\pi^{\pm}$ production from $^3\rm{He}$.}
\label{he3_pion_result_table}
\end{table}

\begin{table*}
\begin{tabular}{|c|c|c|c|}\hline
$<$$p_T$$>$       &      $<$$x_F$$>$      &      $\pi^{+}$       &     $\pi^{-}$                   \\
(GeV/c)           &                       &    ($A_\text{LT}^{\cos(\phi_{s})}\pm$ Stat.$\pm$ Sys.)       &    ($A_\text{LT}^{\cos(\phi_{s})}\pm$ Stat.$\pm$ Sys.)     \\ \hline
0.60              & -0.269                & -0.063$\pm$0.014$\pm$0.012  &   0.024$\pm$0.005$\pm$0.006    \\
0.64              & -0.263                & -0.049$\pm$0.016$\pm$0.011  &  0.020$\pm$0.006$\pm$0.006       \\
0.68              & -0.254                &  -0.032$\pm$0.015$\pm$0.011  &  0.019$\pm$0.005$\pm$0.005    \\   \hline
\end{tabular}
\caption{Tabulated results of $p_{T}$ dependent $A_\text{LT}^{\cos(\phi_{s})}$ for $\pi^{\pm}$ production from the neutron.
A negative $x_{F}$ indicates that the produced hadron is moving backwards with respect to the nucleon momentum direction
in the center-of-mass frame of the e+N system.}
\label{neutron_pion_result_table}
\end{table*}

\begin{figure}
\begin{center}
\includegraphics[width=85mm]{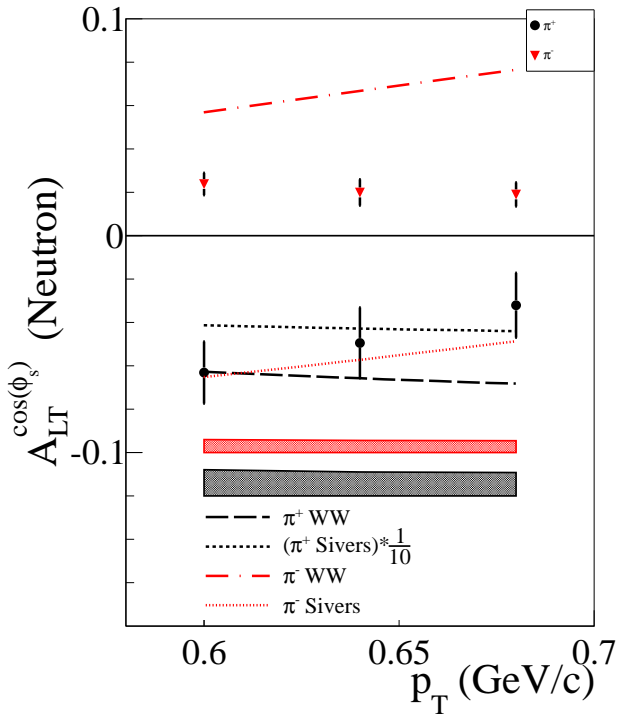}
\end{center}
\caption{(Color online) Beam-target double-spin asymmetries $A_\text{LT}^{\cos(\phi_{s})}$ for $\pi^{\pm}$ production from the neutron as
a function of $p_{T}$. The systematic
uncertainty is shown as a band. The red (top) band is the systematic uncertainty band
for $\pi^{-}$, and the black (bottom) band is the systematic uncertainty band for $\pi^{+}$.
Predictions from collinear factorization by using two different scenarios \cite{Alt_paper}
(Sivers function and Wandzura-Wilczek(WW)-type approximation) 
are shown as well. Please note that the prediction for $\pi^{+}$ by using the Sivers function is scaled by a factor of $\frac{1}{10}$.
}
\label{neutron_pion_result_plot}
\end{figure}

Neutron asymmetries for pion production were obtained from the $^3\rm{He}$ asymmetries using the effective
polarizations of the proton and neutron in polarized $^3\rm{He}$ using the equation \cite{scopetta2007},
\begin{equation}
A^{^{3}\text{He}}_{\text{LT}} =  \text{P}_n(1-f_p){A}^{n}_{\text{LT}} + \text{P}_pf_p{A}^p_{\text{LT}},
\label{eqn_neutron_ext}
\end{equation}
where $A^{^{3}\text{He}}_{\text{LT}}$ is the measured $^3\rm{He}$ asymmetry. P$_n$ =
0.86$^{+0.036}_{-0.02}$ and P$_p$ = -0.028$^{+0.009}_{-0.004}$ are the
effective polarization of the neutron and proton, respectively. The proton 
dilutions, $f_p$ = $\frac{2\sigma_p}{\sigma_{^{3}\rm{He}}}$, in 
$^3\rm{He}$ were measured directly by measuring yields from unpolarized 
hydrogen and $^3\rm{He}$ targets. The average of $f_p$ for 
$\pi^{+}$ was 0.844 $\pm$ 0.007 and for $\pi^{-}$, 0.732 $\pm$ 0.005. 
Since there were no $A_\text{LT}$ experimental data from the proton, and the contribution to the 
final $^3\rm{He}$ asymmetry from polarized protons in polarized $^3\rm{He}$ is small due to the 
small P$_p$, the proton $A_\text{LT}^{p}$ was treated as a systematic uncertainty while 
the neutron asymmetry was extracted from the $^3\rm{He}$ asymmetry.
The beam-target double-spin asymmetry from a polarized proton target was assumed to be no more than $\pm5\%$ 
based on the calculations for a proton target in \cite{Alt_paper}. 
The final $p_{T}$-dependent asymmetries $A_{\text{LT}}^{\cos(\phi_{s})}$ for $\pi^{\pm}$ production from the neutron 
are shown in Figure \ref{neutron_pion_result_plot} and tabulated in Table \ref{neutron_pion_result_table}.
In addition, the kinematic variable $x_{F}$ was also calculated. It is defined as  
$x_{F} = 2p^{CM}/\sqrt{s}$, where $p^{CM}$ is the momentum of the outgoing
hadron along the polarized nucleon's momentum direction in the e+N center-of-mass frame.


\section{Conclusion}

The observed $\pi^{+}$ and $\pi^{-}$ asymmetries from $^3\rm{He}$ and effective 
neutron targets have opposite signs when the target is transversely polarized.
The $\pi^{+}$ and $\pi^{-}$ asymmetries for a vertically polarized target are consistent with zero within the 
experimental uncertainties. 
Although the uncertainty is large, the sign of the $K^{\pm}$ $A_\text{LT}$ is flipped as
the target spin direction is reversed transversely (in the x-direction). 
The $K^{+}$ asymmetry is larger than that of $\pi^{+}$ and they are different in sign. 
If the kaon asymmetry is of partonic origin, it might indicate that sea-quark 
contributions or unfavored fragmentation functions play
a more important role. In addition, higher-order or higher-twist effects might also be possible reasons.
For the proton $A_\text{LT}$, the sign of the asymmetry is flipped as the target spin 
direction is reversed vertically (in the y-direction), while the asymmetry is consistent with zero within the experimental
uncertainty with the target polarized transversely (in the x-direction).
A hypothesis testing was performed to the proton asymmetries, the cos($\phi_s$) dependence 
of the asymmetry can't be excluded within 2-sigma of significance.
One of possible reasons for the interesting behavior of the proton asymmetries might be
that the protons were mostly knocked out from $^3\text{He}$ with nuclear effects. 
In the collinear factorization approximation, $A_\text{LT}$ in inclusive pion production 
was estimated in the JLab 6 GeV kinematic region \cite{Alt_paper}.
The estimations were done using two approximations to calculate the $\widetilde{g}(x)$ while doing 
numerical predictions for $A_\text{LT}$ in inclusive pion production. One is using the 
approximate relation, $\widetilde{g}(x) \approx -f^{\perp}_{1T}(x)$, where $f^{\perp}_{1T}(x)$
is the Sivers function; the other one is using Wandzura-Wilczek(WW)-type approximation,
$\widetilde{g}(x) \approx x \int^{1}_{x} \frac{dy}{y}g_1(y)$.
Calculations based on the two approximations shown 
in Figure \ref{neutron_pion_result_plot} give different predictions. 
Our data are consistent in sign with the prediction using the WW approximation, while the magnitude of the 
predictions is larger than that of our data. 
The calculation by using the Sivers function is not consistent with our data. 
However, one needs to take into account
the current uncertainty of the Sivers function and potential large NLO corrections, which are not included
in the calculation.
We point out that $p_{T}$ in 
our experiment is around 0.64 GeV/c, which is lower than 1 GeV/c where
the theoretical predictions are believed to be reliable. 
In addition, the $A_\text{LT}$ measurements in inclusive hadron production and SIDIS processes are linked 
by the definition of $\widetilde{g}(x)$. 
The behavior of the $\pi^{+}$ and $\pi^{-}$ $A_\text{LT}^{\cos(\phi_{s})}$ 
with opposite sign is similar to that in the SIDIS measurement in \cite{Huang2012},
while the size of the asymmetries in inclusive and SIDIS processes are different.
However, one has to be aware that the 
kinematic coverage for the non-detected electrons in the inclusive hadron production processes
is larger than that of the electrons in the SIDIS processes and the production mechanism can also be different.
To fully interpret the data, one has to understand the mechanism of  
inclusive hadron production in different kinematic regions and the main 
contributions to the double-spin asymmetry.

In summary, we have reported the measurement of $A_\text{LT}$ in the inclusive 
hadron production reaction using longitudinally polarized electrons scattered from 
a transversely polarized $^3\text{He}$ target. Non-zero asymmetries were observed for 
charged pions from a transversely polarized target. The asymmetries in $\pi^{+}$ and $\pi^{-}$ production
have opposite signs. The asymmetries are compared to calculations from collinear factorization,
and the signs of the asymmetries are consistent with calculations using the WW approximation. 
To fully understand inclusive hadron 
production in terms of parton distributions and correlations among partons, new theoretical and 
experimental efforts should be carried out. Future experiments at Jefferson Lab 
\cite{Jlab12GeV, SoLID_white_paper} and a future electron-ion collider (EIC) \cite{EIC_white_paper} 
will extend the measurement to a broad $p_T$ range and a much higher precision. 

\section{Acknowledgements}

We acknowledge the outstanding support of the JLab Hall
A staff and the Accelerator Division in accomplishing
this experiment. This work was supported in part by the
U. S. National Science Foundation, and by Department of Energy (DOE)
contract number DE-AC05-06OR23177, under which the Jefferson Science
Associates operates the Thomas Jefferson National Accelerator
Facility. This work was also supported by the National Natural Science
Foundation of China under Grants No. 11135002 and No. 11120101004 and 
the UK Science and Technology Facilities Council under Grants No. 57071/1 
and No. 50727/1.

\nocite{*}
\bibliography{references}

\begin{thebibliography}{23}%
\makeatletter
\providecommand \@ifxundefined [1]{%
 \@ifx{#1\undefined}
}%
\providecommand \@ifnum [1]{%
 \ifnum #1\expandafter \@firstoftwo
 \else \expandafter \@secondoftwo
 \fi
}%
\providecommand \@ifx [1]{%
 \ifx #1\expandafter \@firstoftwo
 \else \expandafter \@secondoftwo
 \fi
}%
\providecommand \natexlab [1]{#1}%
\providecommand \enquote  [1]{``#1''}%
\providecommand \bibnamefont  [1]{#1}%
\providecommand \bibfnamefont [1]{#1}%
\providecommand \citenamefont [1]{#1}%
\providecommand \href@noop [0]{\@secondoftwo}%
\providecommand \href [0]{\begingroup \@sanitize@url \@href}%
\providecommand \@href[1]{\@@startlink{#1}\@@href}%
\providecommand \@@href[1]{\endgroup#1\@@endlink}%
\providecommand \@sanitize@url [0]{\catcode `\\12\catcode `\$12\catcode
  `\&12\catcode `\#12\catcode `\^12\catcode `\_12\catcode `\%12\relax}%
\providecommand \@@startlink[1]{}%
\providecommand \@@endlink[0]{}%
\providecommand \url  [0]{\begingroup\@sanitize@url \@url }%
\providecommand \@url [1]{\endgroup\@href {#1}{\urlprefix }}%
\providecommand \urlprefix  [0]{URL }%
\providecommand \Eprint [0]{\href }%
\providecommand \doibase [0]{http://dx.doi.org/}%
\providecommand \selectlanguage [0]{\@gobble}%
\providecommand \bibinfo  [0]{\@secondoftwo}%
\providecommand \bibfield  [0]{\@secondoftwo}%
\providecommand \translation [1]{[#1]}%
\providecommand \BibitemOpen [0]{}%
\providecommand \bibitemStop [0]{}%
\providecommand \bibitemNoStop [0]{.\EOS\space}%
\providecommand \EOS [0]{\spacefactor3000\relax}%
\providecommand \BibitemShut  [1]{\csname bibitem#1\endcsname}%
\let\auto@bib@innerbib\@empty
\bibitem [{\citenamefont {Kuhn}\ \emph {et~al.}(2009)\citenamefont {Kuhn},
  \citenamefont {Chen},\ and\ \citenamefont {Leader}}]{JPreview}%
  \BibitemOpen
  \bibfield  {author} {\bibinfo {author} {\bibfnamefont {S.}~\bibnamefont
  {Kuhn}}, \bibinfo {author} {\bibfnamefont {J.-P.}\ \bibnamefont {Chen}}, \
  and\ \bibinfo {author} {\bibfnamefont {E.}~\bibnamefont {Leader}},\ }\href
  {\doibase http://dx.doi.org/10.1016/j.ppnp.2009.02.001} {\bibfield  {journal}
  {\bibinfo  {journal} {Prog. in Part. and Nucl. Phys.}\ }\textbf {\bibinfo
  {volume} {63}},\ \bibinfo {pages} {1 } (\bibinfo {year} {2009})}\BibitemShut
  {NoStop}%
\bibitem [{\citenamefont {Huang}\ \emph {et~al.}(2012)\citenamefont {Huang}
  \emph {et~al.}}]{Huang2012}%
  \BibitemOpen
  \bibfield  {author} {\bibinfo {author} {\bibfnamefont {J.}~\bibnamefont
  {Huang}} \emph {et~al.},\ }\href {\doibase 10.1103/PhysRevLett.108.052001}
  {\bibfield  {journal} {\bibinfo  {journal} {Phys. Rev. Lett.}\ }\textbf
  {\bibinfo {volume} {108}},\ \bibinfo {pages} {052001} (\bibinfo {year}
  {2012})}\BibitemShut {NoStop}%
\bibitem [{\citenamefont {Afanasev}\ \emph {et~al.}(1998)\citenamefont
  {Afanasev}, \citenamefont {Carlson},\ and\ \citenamefont
  {Wahlquist}}]{PhysRevD.58.054007}%
  \BibitemOpen
  \bibfield  {author} {\bibinfo {author} {\bibfnamefont {A.}~\bibnamefont
  {Afanasev}}, \bibinfo {author} {\bibfnamefont {C.~E.}\ \bibnamefont
  {Carlson}}, \ and\ \bibinfo {author} {\bibfnamefont {C.}~\bibnamefont
  {Wahlquist}},\ }\href {\doibase 10.1103/PhysRevD.58.054007} {\bibfield
  {journal} {\bibinfo  {journal} {Phys. Rev. D}\ }\textbf {\bibinfo {volume}
  {58}},\ \bibinfo {pages} {054007} (\bibinfo {year} {1998})}\BibitemShut
  {NoStop}%
\bibitem [{\citenamefont {Afanasev}\ \emph {et~al.}(2000)\citenamefont
  {Afanasev}, \citenamefont {Carlson},\ and\ \citenamefont
  {Wahlquist}}]{PhysRevD.61.034014}%
  \BibitemOpen
  \bibfield  {author} {\bibinfo {author} {\bibfnamefont {A.}~\bibnamefont
  {Afanasev}}, \bibinfo {author} {\bibfnamefont {C.~E.}\ \bibnamefont
  {Carlson}}, \ and\ \bibinfo {author} {\bibfnamefont {C.}~\bibnamefont
  {Wahlquist}},\ }\href {\doibase 10.1103/PhysRevD.61.034014} {\bibfield
  {journal} {\bibinfo  {journal} {Phys. Rev. D}\ }\textbf {\bibinfo {volume}
  {61}},\ \bibinfo {pages} {034014} (\bibinfo {year} {2000})}\BibitemShut
  {NoStop}%
\bibitem [{\citenamefont {Kanazawa}\ \emph {et~al.}(2014)\citenamefont
  {Kanazawa} \emph {et~al.}}]{Alt_paper}%
  \BibitemOpen
  \bibfield  {author} {\bibinfo {author} {\bibfnamefont {K.}~\bibnamefont
  {Kanazawa}} \emph {et~al.},\ }\href {http://arxiv.org/abs/1411.6459}
  {\bibfield  {journal} {\bibinfo  {journal} {arXiv:1411.6459}\ } (\bibinfo
  {year} {2014})}\BibitemShut {NoStop}%
\bibitem [{\citenamefont {Zhou}\ \emph {et~al.}(2010)\citenamefont {Zhou},
  \citenamefont {Yuan},\ and\ \citenamefont {Liang}}]{PhysRevD.81.054008}%
  \BibitemOpen
  \bibfield  {author} {\bibinfo {author} {\bibfnamefont {J.}~\bibnamefont
  {Zhou}}, \bibinfo {author} {\bibfnamefont {F.}~\bibnamefont {Yuan}}, \ and\
  \bibinfo {author} {\bibfnamefont {Z.-T.}\ \bibnamefont {Liang}},\ }\href
  {\doibase 10.1103/PhysRevD.81.054008} {\bibfield  {journal} {\bibinfo
  {journal} {Phys. Rev. D}\ }\textbf {\bibinfo {volume} {81}},\ \bibinfo
  {pages} {054008} (\bibinfo {year} {2010})}\BibitemShut {NoStop}%
\bibitem [{\citenamefont {Kang}\ \emph {et~al.}(2011)\citenamefont {Kang} \emph
  {et~al.}}]{PhysRevD.84.034046}%
  \BibitemOpen
  \bibfield  {author} {\bibinfo {author} {\bibfnamefont {Z.-B.}\ \bibnamefont
  {Kang}} \emph {et~al.},\ }\href {\doibase 10.1103/PhysRevD.84.034046}
  {\bibfield  {journal} {\bibinfo  {journal} {Phys. Rev. D}\ }\textbf {\bibinfo
  {volume} {84}},\ \bibinfo {pages} {034046} (\bibinfo {year}
  {2011})}\BibitemShut {NoStop}%
\bibitem [{\citenamefont {Metz}\ \emph {et~al.}(2012)\citenamefont {Metz},
  \citenamefont {Pitonyak}, \citenamefont {Sch\"afer},\ and\ \citenamefont
  {Zhou}}]{PhysRevD.86.114020}%
  \BibitemOpen
  \bibfield  {author} {\bibinfo {author} {\bibfnamefont {A.}~\bibnamefont
  {Metz}}, \bibinfo {author} {\bibfnamefont {D.}~\bibnamefont {Pitonyak}},
  \bibinfo {author} {\bibfnamefont {A.}~\bibnamefont {Sch\"afer}}, \ and\
  \bibinfo {author} {\bibfnamefont {J.}~\bibnamefont {Zhou}},\ }\href {\doibase
  10.1103/PhysRevD.86.114020} {\bibfield  {journal} {\bibinfo  {journal} {Phys.
  Rev. D}\ }\textbf {\bibinfo {volume} {86}},\ \bibinfo {pages} {114020}
  (\bibinfo {year} {2012})}\BibitemShut {NoStop}%
\bibitem [{\citenamefont {Alcorn}\ \emph {et~al.}(2004)\citenamefont {Alcorn}
  \emph {et~al.}}]{Alcorn2004}%
  \BibitemOpen
  \bibfield  {author} {\bibinfo {author} {\bibfnamefont {J.}~\bibnamefont
  {Alcorn}} \emph {et~al.},\ }\href {\doibase 10.1016/j.nima.2003.11.415}
  {\bibfield  {journal} {\bibinfo  {journal} {Nucl. Instrum. Meth. A}\ }\textbf
  {\bibinfo {volume} {522}},\ \bibinfo {pages} {294 } (\bibinfo {year}
  {2004})}\BibitemShut {NoStop}%
\bibitem [{\citenamefont {Qian}\ \emph {et~al.}(2011)\citenamefont {Qian} \emph
  {et~al.}}]{Qian2011}%
  \BibitemOpen
  \bibfield  {author} {\bibinfo {author} {\bibfnamefont {X.}~\bibnamefont
  {Qian}} \emph {et~al.},\ }\href {\doibase 10.1103/PhysRevLett.107.072003}
  {\bibfield  {journal} {\bibinfo  {journal} {Phys. Rev. Lett.}\ }\textbf
  {\bibinfo {volume} {107}},\ \bibinfo {pages} {072003} (\bibinfo {year}
  {2011})}\BibitemShut {NoStop}%
\bibitem [{\citenamefont {Allada}\ \emph {et~al.}(2014)\citenamefont {Allada},
  \citenamefont {Zhao} \emph {et~al.}}]{PhysRevC.89.042201}%
  \BibitemOpen
  \bibfield  {author} {\bibinfo {author} {\bibfnamefont {K.}~\bibnamefont
  {Allada}}, \bibinfo {author} {\bibfnamefont {Y.~X.}\ \bibnamefont {Zhao}},
  \emph {et~al.} (\bibinfo {collaboration} {Jefferson Lab Hall A
  Collaboration}),\ }\href {\doibase 10.1103/PhysRevC.89.042201} {\bibfield
  {journal} {\bibinfo  {journal} {Phys. Rev. C}\ }\textbf {\bibinfo {volume}
  {89}},\ \bibinfo {pages} {042201} (\bibinfo {year} {2014})}\BibitemShut
  {NoStop}%
\bibitem [{\citenamefont {Zhao}\ \emph {et~al.}(2014)\citenamefont {Zhao} \emph
  {et~al.}}]{PhysRevC.90.055201}%
  \BibitemOpen
  \bibfield  {author} {\bibinfo {author} {\bibfnamefont {Y.~X.}\ \bibnamefont
  {Zhao}} \emph {et~al.} (\bibinfo {collaboration} {Jefferson Lab Hall A
  Collaboration}),\ }\href {\doibase 10.1103/PhysRevC.90.055201} {\bibfield
  {journal} {\bibinfo  {journal} {Phys. Rev. C}\ }\textbf {\bibinfo {volume}
  {90}},\ \bibinfo {pages} {055201} (\bibinfo {year} {2014})}\BibitemShut
  {NoStop}%
\bibitem [{\citenamefont {Sinclair}\ \emph {et~al.}(2007)\citenamefont
  {Sinclair} \emph {et~al.}}]{PhysRevSTAB.10.023501}%
  \BibitemOpen
  \bibfield  {author} {\bibinfo {author} {\bibfnamefont {C.~K.}\ \bibnamefont
  {Sinclair}} \emph {et~al.},\ }\href {\doibase 10.1103/PhysRevSTAB.10.023501}
  {\bibfield  {journal} {\bibinfo  {journal} {Phys. Rev. ST Accel. Beams}\
  }\textbf {\bibinfo {volume} {10}},\ \bibinfo {pages} {023501} (\bibinfo
  {year} {2007})}\BibitemShut {NoStop}%
\bibitem [{\citenamefont {Androić}\ \emph {et~al.}(2011)\citenamefont
  {Androić} \emph {et~al.}}]{Androić201159}%
  \BibitemOpen
  \bibfield  {author} {\bibinfo {author} {\bibfnamefont {D.}~\bibnamefont
  {Androić}} \emph {et~al.},\ }\href {\doibase
  http://dx.doi.org/10.1016/j.nima.2011.04.031} {\bibfield  {journal} {\bibinfo
   {journal} {Nucl. Instrum. Meth. A}\ }\textbf {\bibinfo {volume} {646}},\
  \bibinfo {pages} {59 } (\bibinfo {year} {2011})}\BibitemShut {NoStop}%
\bibitem [{\citenamefont {Bissey}\ \emph {et~al.}(2002)\citenamefont {Bissey},
  \citenamefont {Guzey}, \citenamefont {Strikman},\ and\ \citenamefont
  {Thomas}}]{Bissey2001}%
  \BibitemOpen
  \bibfield  {author} {\bibinfo {author} {\bibfnamefont {F.}~\bibnamefont
  {Bissey}}, \bibinfo {author} {\bibfnamefont {V.}~\bibnamefont {Guzey}},
  \bibinfo {author} {\bibfnamefont {M.}~\bibnamefont {Strikman}}, \ and\
  \bibinfo {author} {\bibfnamefont {A.}~\bibnamefont {Thomas}},\ }\href
  {\doibase 10.1103/PhysRevC.65.064317} {\bibfield  {journal} {\bibinfo
  {journal} {Phys. Rev. C}\ }\textbf {\bibinfo {volume} {65}},\ \bibinfo
  {pages} {064317} (\bibinfo {year} {2002})}\BibitemShut {NoStop}%
\bibitem [{\citenamefont {Babcock}\ \emph {et~al.}(2005)\citenamefont
  {Babcock}, \citenamefont {Nelson}, \citenamefont {Kadlecek},\ and\
  \citenamefont {Walker}}]{Babcock2005}%
  \BibitemOpen
  \bibfield  {author} {\bibinfo {author} {\bibfnamefont {E.}~\bibnamefont
  {Babcock}}, \bibinfo {author} {\bibfnamefont {I.~A.}\ \bibnamefont {Nelson}},
  \bibinfo {author} {\bibfnamefont {S.}~\bibnamefont {Kadlecek}}, \ and\
  \bibinfo {author} {\bibfnamefont {T.~G.}\ \bibnamefont {Walker}},\ }\href
  {\doibase 10.1103/PhysRevA.71.013414} {\bibfield  {journal} {\bibinfo
  {journal} {Phys. Rev. A}\ }\textbf {\bibinfo {volume} {71}},\ \bibinfo
  {pages} {013414} (\bibinfo {year} {2005})}\BibitemShut {NoStop}%
\bibitem [{\citenamefont {Singh}\ \emph {et~al.}(2013)\citenamefont {Singh}
  \emph {et~al.}}]{E06010_target}%
  \BibitemOpen
  \bibfield  {author} {\bibinfo {author} {\bibfnamefont {J.}~\bibnamefont
  {Singh}} \emph {et~al.},\ }\href {http://arxiv.org/abs/1309.4004} {\bibfield
  {journal} {\bibinfo  {journal} {arXiv:1309.4004}\ } (\bibinfo {year}
  {2013})}\BibitemShut {NoStop}%
\bibitem [{\citenamefont {Romalis}\ and\ \citenamefont {Cates}(1998)}]{EPR}%
  \BibitemOpen
  \bibfield  {author} {\bibinfo {author} {\bibfnamefont {M.~V.}\ \bibnamefont
  {Romalis}}\ and\ \bibinfo {author} {\bibfnamefont {G.~D.}\ \bibnamefont
  {Cates}},\ }\href {\doibase 10.1103/PhysRevA.58.3004} {\bibfield  {journal}
  {\bibinfo  {journal} {Phys. Rev. A}\ }\textbf {\bibinfo {volume} {58}},\
  \bibinfo {pages} {3004} (\bibinfo {year} {1998})}\BibitemShut {NoStop}%
\bibitem [{\citenamefont {Wang}(2011)}]{youcai_thesis}%
  \BibitemOpen
  \bibfield  {author} {\bibinfo {author} {\bibfnamefont {Y.}~\bibnamefont
  {Wang}},\ }\href {http://hallaweb.jlab.org/experiment/transversity/thesis}
  {Ph.D. thesis},\ \bibinfo  {school} {UIUC} (\bibinfo {year}
  {2011})\BibitemShut {NoStop}%
\bibitem [{\citenamefont {Scopetta}(2007)}]{scopetta2007}%
  \BibitemOpen
  \bibfield  {author} {\bibinfo {author} {\bibfnamefont {S.}~\bibnamefont
  {Scopetta}},\ }\href {\doibase 10.1103/PhysRevD.75.054005} {\bibfield
  {journal} {\bibinfo  {journal} {Phys. Rev. D}\ }\textbf {\bibinfo {volume}
  {75}},\ \bibinfo {pages} {054005} (\bibinfo {year} {2007})}\BibitemShut
  {NoStop}%
\bibitem [{\citenamefont {Gao}\ \emph {et~al.}(2011)\citenamefont {Gao} \emph
  {et~al.}}]{Jlab12GeV}%
  \BibitemOpen
  \bibfield  {author} {\bibinfo {author} {\bibfnamefont {H.}~\bibnamefont
  {Gao}} \emph {et~al.},\ }\href {\doibase 10.1140/epjp/i2011-11002-4}
  {\bibfield  {journal} {\bibinfo  {journal} {Eur. Phys. J.}\ }\textbf
  {\bibinfo {volume} {126}},\ \bibinfo {pages} {1} (\bibinfo {year}
  {2011})}\BibitemShut {NoStop}%
\bibitem [{\citenamefont {Chen}\ \emph {et~al.}(2014)\citenamefont {Chen} \emph
  {et~al.}}]{SoLID_white_paper}%
  \BibitemOpen
  \bibfield  {author} {\bibinfo {author} {\bibfnamefont {J.-P.}\ \bibnamefont
  {Chen}} \emph {et~al.},\ }\href {http://arxiv.org/abs/1409.7741} {\bibfield
  {journal} {\bibinfo  {journal} {arXiv:1409.7741}\ } (\bibinfo {year}
  {2014})}\BibitemShut {NoStop}%
\bibitem [{\citenamefont {Accardi}\ \emph {et~al.}(2014)\citenamefont {Accardi}
  \emph {et~al.}}]{EIC_white_paper}%
  \BibitemOpen
  \bibfield  {author} {\bibinfo {author} {\bibfnamefont {A.}~\bibnamefont
  {Accardi}} \emph {et~al.},\ }\href {http://arxiv.org/abs/1212.1701}
  {\bibfield  {journal} {\bibinfo  {journal} {arXiv:1212.1701}\ } (\bibinfo
  {year} {2014})}\BibitemShut {NoStop}%
\end{thebibliography}%

\begin{table*}
\begin{center}
\begin{tabular}{|c|c|c|} \hline
    &     $\phi_{s}=90^{o}$        &    $\phi_{s}=270^{o}$            \\  \hline
$\pi^{+}$   &  0.0001$\pm$0.0015$\pm$0.0008 & 0.0006$\pm$0.0015$\pm$0.0008    \\  \hline
$\pi^{-}$   & -0.0007$\pm$0.0011$\pm$0.0008  & -0.0004$\pm$0.0011$\pm$0.0008 \\  \hline
$K^{+}$     & 0.007$\pm$0.023$\pm$0.003      & 0.017$\pm$0.024$\pm$0.004        \\  \hline
$K^{-}$     & 0.011$\pm$0.032$\pm$0.005      & 0.044$\pm$0.033$\pm$0.006        \\  \hline
P           & 0.0073$\pm$0.0047$\pm$0.0009    & -0.0083$\pm$0.0047$\pm$0.001    \\  \hline \hline
            &       $\phi_{s}=0^{o}$         &   $\phi_{s}=180^{o}$      \\  \hline
$\pi^{+}$   &   -0.0058$\pm$0.0016$\pm$0.0009 & 0.0071$\pm$0.0016$\pm$0.0009   \\  \hline
$\pi^{-}$   &  0.0044$\pm$0.0011$\pm$0.0008  & -0.0056$\pm$0.0011$\pm$0.0009 \\  \hline
$K^{+}$     &   0.055$\pm$0.030$\pm$0.006    & -0.066$\pm$0.031$\pm$0.006    \\  \hline
$K^{-}$     &  0.014$\pm$0.040$\pm$0.006      & -0.045$\pm$0.041$\pm$0.007     \\  \hline
P           &  0.0032$\pm$0.0067$\pm$0.0008  & -0.0013$\pm$0.0069$\pm$0.0008  \\  \hline
\end{tabular}
\caption{ The beam-target double-spin asymmetries $A_\text{LT}$ from $^3\rm{He}$ for different hadron species. The data structure
follows the format of asymmetry$\pm$statistical uncertainty$\pm$systematic uncertainty. The average $p_{T}$ is 0.64 GeV/c.}
\label{he3_result_all_species_table}
\end{center}
\end{table*}

\end{document}